# Neutron stars: probes for high density baryon pairing [1] [2]

Dany PAGE

*Instituto de Astronomía, UNAM*

*Circuito Exterior, C.U., Apartado Postal 70-264, 04510 México D.F., México*

ABSTRACT. We consider, in general terms, the early thermal evolution of an isolated neutron star, i.e., during the first $10^5$ years after the supernova explosion when the cooling is driven by neutrino emission from the core. It is shown that, if fast neutrino emission is occuring, the surface temperature is actually determined by the critical temperature $T_c$ at which the core baryons become superfluid, and almost nothing else. In particular, the dependence on the actual neutrino emission process is very weak. In other terms, fast neutrino cooling neutron stars are thermometers for the highest $T_c$ superfluid (superconductor) in the Universe. The observational data are briefly presented.

RESUMEN. Se considera, en términos generales, la evolución térmica de una estrella de neutrones joven, i.e., durante los primeros $10^5$ años después de la explosión de una supernova, cuando el enfriamiento es dominado por la emisión de neutrinos en el interior. Se muestra que, cuando occure emisión rápida de neutrinos, la temperatura superficial está determinada casi totalmente por la temperatura crítica $T_c$ a la cual los bariones del interior se vuelven superfluidos. Esto significa, en particular, que la temperatura depende muy poco del tipo de emisión rápida. En otras palabras, las estrellas de neutrones con enfriamiento rápido son termómetros para el superfluido (superconductor) de más alta $T_c$ en el Universo. Los datos observacionales se presentan brevemente.

1. INTRODUCTION

Due to the extreme conditions present in their interiors, neutron stars (NS) are unique laboratories for nuclear physics. A most promising tool for studying the NS interiors is the modeling of their thermal evolution which, at early stages, is driven by neutrino emission from matter at supranuclear densities. We summarize in this paper the basic concepts necessary for understanding and studying fast neutrino cooling of neutron stars, showing what can be learned from this study and underlining its limitations.

The theory of neutron star cooling started more than 30 years ago and often progressed at a slow pace due to the paucity of data, but the recent launch of the *ROSAT* observatory revived the field. Deep *ROSAT* observations have finally provided strong evidence for detection of surface thermal radiation from four nearby neutron stars [1]. The interpretation of these data has already provided new insights into the structure of the neutron star's interior [2] and surface [3, 4]. However, the results presented here seem to require improved data or deeper analysis of the available ones.

---

[1] Work presented at the **XVIII Nuclear Physics Symposium at Oaxtepec**, Oaxtepec, México, Jan. 4-7, 1995.

[2] This work was supported by a UNAM-DGAPA grant IN105794.

*1.1 'Neutron' Stars ?*

Two important questions that can be asked about matter at supranuclear density are the 'chemical' composition and the state of this matter, superfluid or normal. At densities around and slightly above the nuclear density, $\rho_0 = 2.8 \cdot 10^{14}$ g cm$^{-3}$, matter is made mostly of neutrons with a small admixture of protons and electrons (and muons) to maintain charge neutrality. However, at densities between $2 - 4\rho_0$ new particles are expected to appear: pions, kaons, hyperons, or simply a large fraction of protons. Who will come up first and at which critical density is still a question of debate, but there is now little doubt that the term 'neutron star' is not representative of the real content of these objects. Pion condensation is presently considered as improbable and has been replaced by kaon condensation while the appearance of quark matter is still a question of personal taste. Hyperons are serious candidates but the present meagerness of our knowledge of hyperon-hyperon interactions precludes any convincing study. As for the concentration of protons, in absence of non nucleonic particles, it is determined by the symmetry energy of dense matter, also a poorly know quantity.

All these possibilities share a common property: they increase enormously the neutrino emission compared to the slow modified Urca process [5] relevant within the conservative view of neutron stars ('standard model') These fast neutrino emission mechanisms overwhelm the standard mechanism by 3 to 6 orders of magnitude at a temperature of $10^9$ K: the occurrence of any of these processes will of course have a dramatic effect on the cooling of a neutron star. We refer to the reviews [6] for a detailed discussion. The study of neutron star cooling offers thus the possibility to find observational evidence for the presence of these 'exotic' states of matter. The cooling of neutron stars under the 'standard model' has been presented in detail in the classic paper of Nomoto & Tsuruta [7].

*1.2 Baryon pairing*

There must be an undemonstrated fundamental theorem of physics stating that any many-body system must be solid or superfluid at low enough temperature. Solidification of neutron matter was popular in the early seventies but has been disproved since then: the alternative is thus superfluidity. Neutron star matter is actually the most obvious place to look at for fermion pairing *à la* BCS: high degeneracy and attractive interactions are naturally present. Nucleons are expected to pair in the $^1S_0$ partial wave at low Fermi momentum and in the $^3P_2$ partial wave at larger momentum. $^1D_2$ pairing at still higher momentum is possible. Thus the free neutrons present in the neutron star's inner crust are paired in the $^1S_0$ state as well as the protons in the core while the core neutrons are paired in the $^3P_2$ state and higher order pairing may happen in the denser regions. However, the calculation of the pairing critical temperature $T_c$ is a famously difficult problem, and published calculations show a very large range of values of $T_c$, particularly for $^3P_2$ neutron pairing. Presently, the only correct attitude toward baryon pairing at high density is, unfortunately, to consider $T_c$ as a free parameter [2].

Pairing will suppress all processes involving excitations. This applies in particular to the specific heat and the neutrino emissivity and will affect dramatically the cooling of neutron stars. These effects where discussed in [8] and [9] for the neutrino emission suppression and in [2] for the specific heat. We will show here that, if we assume that any fast neutrino emission mechanism is occuring, the thermal evolution of a young neutron star is then entirely controlled by baryon pairing.

2. INPUT PHYSICS

When seeing the star from outside what matters is the neutrino emission rate, not the actual mechanism. We will thus write it as

$$\epsilon_\nu = 10^{\mathcal{N}} \cdot T_9^6 \quad \text{erg cm}^{-3} \text{ sec}^{-1}. \tag{1}$$

where the exponent $\mathcal{N}$ takes values between 24 and 27 to mimic the various fast emission processes [6] and $T_9$ is the temperature in units of $10^9$ K. In contrast, the modified Urca process and the other similar processes of the 'standard model' have an emissivity of the order of $10^{19-21} \cdot T_9^8$ erg cm$^{-3}$ sec$^{-1}$. The change in the $T$ dependence comes from the fact that the fast processes involve only 3 degenerate fermions instead of 5: the difference in efficiency is mostly an effect of phase space limitation. We will also consider the critical density $\rho_{crit}$ above which fast emission is allowed as a free parameter.

The superfluidity suppression of the neutrino emissivity $\epsilon_\nu$ and the specific heat $c_v$ is taken into account by the introduction of suppression factors: $\epsilon_\nu \to \mathcal{R}_\nu \epsilon_\nu$ and $c_v \to \mathcal{R}_c c_v$. The $\mathcal{R}_\nu$ and $\mathcal{R}_c$ are functions of $T/T_c$ and depend also on the type of pairing [10]; at $T \ll T_c$ they are proportional to $\exp(-aT_c/T)$, where $a \sim 2-8$, for nodeless gaps, and proportional to $(T/T_c)^2$ for gaps with nodes. For $^1S_0$ pairing the gap is spherically symmetric and nodeless. For $^3P_2$ pairing there are many possible states from which the two apparently most probable have gaps proportional to $(1 + 3\cos^2\theta)$ and $3\sin^2\theta$; we will consider here only nodeless gaps.

The neutrons in the inner crust will always be assumed to be paired in $^1S_0$ wave with the critical temperature calculated in [11]. As for pairing in the core, for both neutrons and protons, we will consider $T_c$ as a free parameter and take it as density independent for a first step. All our calculations use the classic FP equation of state [12]. The other ingredients of our calculations are standard and are described in [8] and [9]. It must be mentioned that all surface temperatures, in Fig. 1 and 2, below $3 \cdot 10^5$ K have to be seen only as illustrative since several important low temperature effects in the envelope have not been included. Magnetic field effects in the envelope have also been neglected.

3. RESULTS AND DISCUSSION

The main points we will consider here are the effects of the actual rate of neutrino emission, Equ. 1, the critical density $\rho_{crit}$ for the onset of fast neutrino emission and the critical temperature $T_c$ of the pairing phase transition.

We show in Fig. 1 cooling curves for various values of $\rho_{crit}$, $\mathcal{N}$ and $T_c$ for core neutron pairing with a nodeless gap. For the first ten years, approximately, the surface temperature $T_S$ is independent of the core evolution: this is due to the finite diffusion time through the crust heat blanket and the outer part of the core [13]. Once the cold wave from the core reaches the surface, the temperature drops precipitously: a higher $\rho_{crit}$ naturally results in a delay in this temperature drop (thicker blanket). After this, the cooling is much

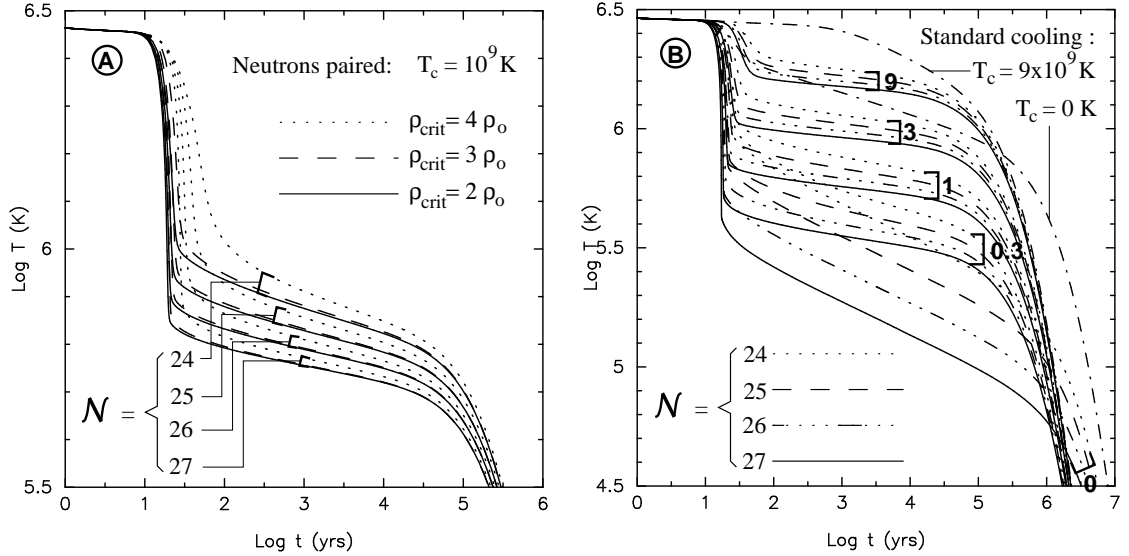

Figure 1: Cooling with fast neutrino emission and neutron pairing. Core neutrons are paired in $^3P_2$ state with a nodeless gap. The curves show the evolution of the surface temperature $T_S$ as function of the neutron star age. **A**: the three sets of four curves each show the spread in $T_S$ due to the change in the fast neutrino emission efficiency (Equ. 1) and the critical density $\rho_{crit}$ for onset of fast neutrino emission. **B**: the five sets of four curves each show the strong dependence of $T_S$ on the critical temperature $T_c$. The values of $T_c$ (in units of $10^9$ K) from 9 down to 0.3 and 0 (i.e. no pairing) are indicated next to each quadruplet of curves; curves within a given quadruplet differ by the value of $\mathcal{N}$. The two curves labeled 'standard cooling' show, for comparison, the cooling in absence of fast neutrino emission with and without core neutron pairing. During the photon cooling era ($t \gtrsim 10^5$ yrs) all models with neutron pairing (including the 'standard model') approach the same asymptotic trajectory since they have the same specific heat, and models without pairing ($T_c = 0$) cool more slowly due to their larger specific heat.

slower for about $10^5$ years until photon cooling from the surface emission takes over and the (negative) slope of the curves decreases strongly. From now on we will concentrate on the evolution at ages between 1,000 and 50,000 years. Figure 1 A shows clearly that, when fast neutrino emission is stopped by pairing, the actual size of the fast neutrino emitting 'pit' is of little importance, unless $\rho_{crit}$ were very close to the central density and the pit extremely small. The masses and radii of the pits, in our models, are (1.04 $M_\odot$, 8.4 km) for $\rho_{crit} = 2\rho_0$, (0.63 $M_\odot$, 6.7 km) for $\rho_{crit} = 3\rho_0$, and (0.15 $M_\odot$, 3.9 km) for $\rho_{crit} = 4\rho_0$. The star has a mass of 1.4 $M_\odot$ and a radius of 10.7 km. Even the smallest pit of 0.15 $M_\odot$ results in a $T_S$ that is observationally indistinguishable from the larger ones. Had we used a different EOS, which means a different central density and pit size, the resulting surface temperature would be the same, within observational uncertainties. The same Fig. 1 A also shows the small dependence on the actual rate of fast neutrino emission:

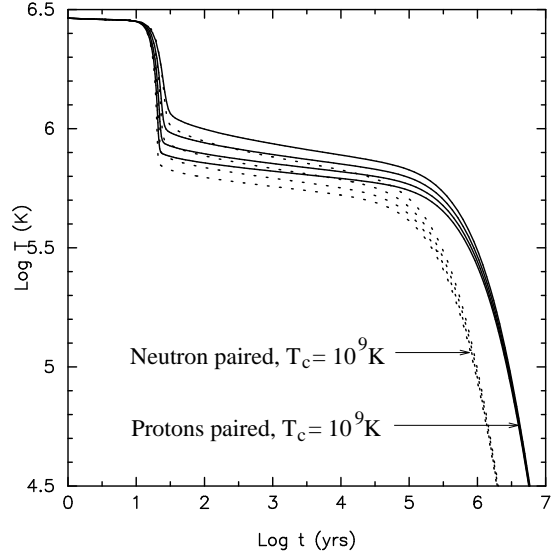

Figure 2: Comparison of fast cooling with neutron pairing (nodeless gaps) and proton pairing. The four curves in each case correspond to $\mathcal{N}$ =24, 25, 26 and 27 by order of decreasing temperature. For a given $T_c$, the neutron gap $\Delta_n$ is larger than the proton gap $\Delta_p$ and the effect of the early neutrino emission suppression thus more efficient, resulting in higher surface temperatures till ages around $10^5$ yrs. During the photon cooling era ($t \gtrsim 10^5$ yrs) neutron stars with neutrons paired cool much faster since they have a lower specific heat: about 75% of the star's specific heat is due to the neutrons and 20% to the protons [2].

increasing $\epsilon_\nu$ by *four* orders of magnitude affects $T_s$ by at most 50% for $T_c = 10^9$ K. The reason for this very weak dependence on $\mathcal{N}$ is simple: the core temperature $T$ drops very fast till $T \lesssim T_c$ and then the neutrino emission is progressively turned off. With the various fast neutrino emission processes, $T$ will reach $T_c$ at different times but always very early on, while if only slow neutrino emission is active $T_c$ is reached much later and the resulting evolution is qualitatively different. The determinant parameter which controls the thermal evolution is thus the critical temperature $T_c$ for the onset of core neutron pairing as is clearly demonstrated in Fig. 1 B: for the range of variation of $T_c$ considered there, $T_S$ changes by an order of magnitude, i.e., the luminosity changes by four orders of magnitude. In the case of fast neutrino emission by the direct Urca process this result was already clearly demonstrated in [9]. This strong dependence of $T_S$ on $T_c$ precludes us of predicting the thermal evolution of fast cooling neutron stars because of the large uncertainties in the theoretical calculations of $T_c$. Page & Applegate [9] proposed to invert the problem: *if we can measure the surface temperature of a fast cooling neutron star at ages between* $10^3$ *and* $5 \cdot 10^5$ *years we are actually measuring the critical temperature for core baryon pairing. Figure 1 B shows that this proposition holds independently of the actual fast cooling mechanism.* It appears also that this 'measurement' will be more accurate for high values of $T_c$ ($\lesssim 10^{10}$ K) than for low ones ($\sim 3 \cdot 10^8$ K) where cooling curves become close to the unpaired case. Notice finally that with high values of $T_c$ fast cooling become similar to the 'standard cooling'.

The above models only considered core neutron pairing with a nodeless gap. However, if the neutron gap has nodes the suppression is much less effective and will alter our conclusions. The superfluid's free energy is minimized by a nodeless gap for isotropic homogeneous neutron matter but rotation and/or magnetic field may favor a gap with nodes [14]. This case will be considered in a later paper. For illustration we will consider here the effect of (core) proton pairing, in $^1S_0$ wave, which is also nodeless but with a different relationship between $T_c$ and the gap size $\Delta$. As can be seen in Fig. 2, the difference in the surface temperatures, for the same value of $T_c$, between gapless neutron and (gapless) proton pairing is not very large, at the ages we consider, but sufficient to make the 'measurement' of $T_c$ from observations of $T_S$ less accurate since we do not know *a priori* if the suppression is due to neutron or proton pairing. A detailed study of this combined effect and its use to measure $T_c$ has been presented in [15] in the case of direct Urca cooling.

We have considered here $T_c$ as density independent, a gross approximation. However our results will still be valid in case of density dependent gaps: the cooling will simply be controlled by the region of the pit with the lowest $T_c$. If $\Lambda$ and $\Sigma^-$ hyperons are present they undergo a hyperonic direct Urca process and hyperon pairing would have the same effect as nucleon pairing. Finally, dissipative mechanisms can convert rotational energy from the spin-down into heat and slightly rise the temperatures shown in Fig. 1 and 2 [16], adding another uncertainty to our $T_S - T_c$ relationship.

## 4. The observational data

Evidence for the occurrence of fast neutrino emission will come from the observation of a neutron star with a surface temperature below the lowest prediction of the standard model; this requires an age of less than about $5 \cdot 10^5$ yrs since at later times slow and fast cooling trajectories overlap, see Fig. 1 A and [2]. Unfortunately there are very few candidates for such observations [4] and in only four cases do we have good evidence that surface thermal emission has been detected [1]. Of these four objects only the youngest one, the Vela pulsar of age $\sim 10^4$ yrs, is relevant. Its estimated surface temperature is of the order of $1.5 \cdot 10^6$ K and is marginally inconsistent with standard cooling models: it may be a case for the occurrence of fast neutrino emission but no strong conclusion can be drawn presently. The second oldest, PSR 0656+14, is a limiting case and its temperature estimate can also be accommodated within the standard model due to its age of about $10^5$ yrs, while the other two (PSRs 0630+178 and 1055-52) are definitely too old [2]. Several young pulsars have been detected in the soft X-ray band but, due to their distance, it is not yet possible to separate the surface thermal radiation from the emission of the magnetosphere and the surrounding synchrotron nebula, and only upper limits for their $T_S$ have been obtained, none of them providing a strong case for fast cooling. Better analysis of the Vela data may provide us with the first serious case, otherwise we will have to wait for the next generation of X-ray observatories, to be launched at the end of this decade.

5. CONCLUSIONS

We have shown that the early thermal evolution of neutron stars undergoing fast neutrino cooling is actually determined entirely by the critical temperature $T_c$ of the baryon pairing phase transition within the fast neutrino emitting pit. The actual size of the pit, determined by the EOS and $\rho_{crit}$, is of little importance as well as the actual rate of neutrino emission as long as it is fast. The type of pairing, $^1S_0$ or $^3P_2$ with/without nodes in the gap, is however probably more important: we have considered here only nodeless gaps and gaps with nodes will be studied in a later paper. We can thus extend the proposition of [9], which was restricted to the direct Urca cooling, to all fast neutrino emission mechanisms: *fast neutrino cooling neutron stars are thermometers for the highest $T_c$ superfluid (superconductor) in the Universe.*